\documentclass[aps,prapplied,reprint,superscriptaddress,longbibliography]{revtex4-2}
\usepackage{amsmath,amssymb,graphicx,booktabs}
\usepackage{xcolor}
\usepackage{hyperref}
\hypersetup{colorlinks=true,linkcolor=blue,citecolor=blue,urlcolor=blue}
\newcommand{\avg}[1]{\left\langle #1\right\rangle}
\begin{document}

\newcommand{\LF}[1]{\textcolor{magenta}{#1}}

\title{Optimal quasi-phase matching design for cascaded frequency conversion}
\author{Linran Fan}
\email{linran.fan@utexas.edu}
\affiliation{Chandra Family Department of Electrical and Computer Engineering, The University of Texas at Austin, Austin, Texas 78758, USA}

\begin{abstract}
Cascaded second-order nonlinear interactions enable high-harmonic generation and quantum frequency conversion (QFC) between widely separated wavelengths. Their efficiency depends on how quasi-phase matching (QPM) distributes nonlinear coupling between generating and converting the intermediate field. Conventional designs use two sequential QPM sections, one for each nonlinear process. Here, we derive a spatially weighted binary QPM grating that tracks the evolution of the intermediate field and attains a proven upper bound on the cascaded efficiency in the undepleted, slow-varying model. The optimum efficiency is 35.6\% higher than that of an optimized two-section design, independent of the individual nonlinear efficiencies. Enforcing a minimum domain width of 20\% of the carrier period retains at least 98\% of the weak-conversion output for period ratios between 0.5 and 2. 
We further evaluate low-power optimal cascaded QPM for complete cascaded QFC beyond the undepleted regime, and show that the performance lies within 8\% of a numerically optimized design. 
These results provide a design rule and a performance bound for cascaded frequency conversion, with applications in short-wavelength light generation and low-noise quantum frequency conversion.
\end{abstract}
\maketitle

\section{Introduction}
Efficient frequency conversion is needed to connect optical sources, quantum emitters, and detection channels that operate at different wavelengths~\cite{Kumar1990,Zaske2012,Pelc2011}. Cascaded second-order nonlinear interactions extend the accessible frequency range by coupling the input and output through an intermediate field~\cite{Porat2012,Hannegan2021}. Second-harmonic generation (SHG) followed by sum-frequency generation (SFG), for example, converts a fundamental wave to its third harmonic~\cite{Zhu1997}. Two successive sum- or difference-frequency generation processes can connect widely separated input and output frequencies for quantum frequency conversion (QFC). Splitting the frequency shift into two steps allows pump wavelengths to be chosen to suppress spontaneous parametric down-conversion into the detection band, as demonstrated in low-noise conversion of nitrogen-vacancy-center fluorescence to the telecom C band~\cite{Chamier2025}.

Quasi-phase matching (QPM) compensates the wavevector mismatch through periodic inversion of the nonlinear susceptibility~\cite{Fejer1992}. A straightforward QPM design for cascaded nonlinear processes places two periodically quasi-phase-matched sections in series: the first generates the intermediate field, and the second converts it to the output. 
This design ignores the continuous evolution of the intermediate field: the second process is idle where the intermediate field is generated, and the first is idle where it is converted.

Several approaches go beyond two sections. Quasi-periodic and two-dimensional nonlinear photonic crystals supply multiple reciprocal vectors in a single structure~\cite{Zhu1997,Berger1998,Lifshitz2005},
and aperiodic gratings found by numerical optimization further improve the cascaded efficiency~\cite{Hsu2019}.
Adiabatic designs inspired by stimulated Raman adiabatic passage can achieve efficient cascaded conversion while keeping the intermediate field weak~\cite{Longhi2007,Porat2012,Suchowski2014}.
However, the maximum cascaded efficiency achievable with a binary poling pattern, and the pattern that reaches it, have not been established.

In this work, we derive the optimum binary QPM pattern for cascaded conversion in the undepleted regime and assess its pump-power requirement and fabrication tolerance. Comparing two-section QPM with a uniform beating grating motivates a weighting envelope that favors intermediate-field generation near the input and its conversion near the output. This envelope attains a proven upper bound of the slow-varying model and gives 35.6\% higher weak-conversion output than optimized two-section QPM. We then evaluate the penalty imposed by a finite minimum domain width. Finally, we show that for complete QFC in a representative thin-film lithium niobate waveguide, the same envelope requires less pump power than optimized sections.

\section{Coupled-wave model and effective nonlinear coupling}
We first consider a waveguide that generates a third harmonic through second-harmonic generation (SHG) followed by sum-frequency generation (SFG). Let $A_\omega$, $A_{2\omega}$, and $A_{3\omega}$ be complex modal amplitudes of the fundamental, second-harmonic, and third-harmonic fields, respectively. We consider negligible propagation loss and weak conversion, so the incident fundamental field is undepleted. With no injected second or third harmonic, the coupled-wave equations are
\begin{equation}
\begin{aligned}
 \frac{dA_{2\omega}}{dz}&=i\frac{\pi}{2}\sqrt{J_1}s(z)e^{-iK_1z}A_\omega^2,\\
 \frac{dA_{3\omega}}{dz}&=i\frac{\pi}{2}\sqrt{J_2}s(z)e^{-iK_2z}A_\omega A_{2\omega}.
\end{aligned}
\label{eq:barefields}
\end{equation}
Here $K_1=k_{2\omega}-2k_\omega$ and $K_2=k_{3\omega}-k_{2\omega}-k_\omega$ are the intrinsic wavevector mismatches, $s(z)=\pm1$ specifies the domain orientation, and $J_1$ and $J_2$ are the power-normalized efficiencies that include the nonlinear susceptibility, transverse mode overlap, and standard first-order QPM factor $2/\pi$; the prefactor $\pi/2$ above therefore gives unit effective coupling for a dedicated periodic grating. 
Periodic domain inversion, described as $s(z)=\operatorname{sgn}[\sin(K_jz)]$, supplies a spatial Fourier component that compensates a phase mismatch, leading to the coherent growth of nonlinear signals. 

For non-periodic poling, we describe the poling pattern through two local QPM coupling factors, $r_1(z)$ and $r_2(z)$
\begin{equation}
 r_j(z)=\frac{\pi}{2}\avg{s\sin(K_j z)},\qquad j=1,2,
 \label{eq:rdefinition}
\end{equation}
where the brackets denote a local average. We assume that $r_j(z)$ varies much more slowly than the rapid domain inversions.
A value $r_j=1$ gives the nonlinear strength of standard periodic QPM, whereas $r_j=0$ gives zero nonlinearity to process $j$. 
The efficiency of an isolated nonlinear process scales as $r_j^2$ at low power.


Averaging Eq.~\eqref{eq:barefields} gives
\begin{equation}
 \frac{dA_{2\omega}}{dz}=\sqrt{J_1}\,r_1 A_\omega^2,
 \qquad
 \frac{dA_{3\omega}}{dz}=\sqrt{J_2}\,r_2 A_\omega A_{2\omega}.
 \label{eq:shgfields}
\end{equation}

Direct integration of Eqs.~\eqref{eq:shgfields} gives the output fields:
\begin{equation}
\begin{aligned}
 A_{3\omega}(L)&=\sqrt{J_1J_2}\,A_\omega^3
 \int_0^L\!dz\,r_2(z)\int_0^z\!dt\,r_1(t).
\end{aligned}
\label{eq:ordered}
\end{equation}
The inner integral accounts for intermediate-field generation upstream of $z$; the outer integral sums its conversion to the output. Denoting this double integral by $\mathcal I$, the power-conversion efficiency is
\begin{align}
 &\eta_{\mathrm{pow}}=\frac{|A_{3\omega}(L)|^2}{|A_\omega|^2}=J_1J_2|A_\omega|^4\,|\mathcal I|^2,\label{eq:eff}\\
 &\mathcal I=\int_0^L\!dz\,r_2(z)\int_0^z\!dt\,r_1(t).\label{eq:cascaded_integral}
\end{align}


For two-step QFC, let $A_s,A_i,A_o$ denote the input, intermediate, and output signal amplitudes. Under undepleted pumps of powers $P_{p1},P_{p2}$ and negligible signal depletion, the corresponding equations are
\begin{equation}
 \frac{dA_i}{dz}=\sqrt{J_1P_{p1}}\,r_1 A_s,
 \qquad
 \frac{dA_o}{dz}=\sqrt{J_2P_{p2}}\,r_2 A_i.
 \label{eq:qfcweakfields}
\end{equation}
The output signal is calculated as 
\begin{equation}
    A_o(L)=\sqrt{J_1J_2P_{p1}P_{p2}}\,A_s
 \int_0^L\!dz\,r_2(z)\int_0^z\!dt\,r_1(t).
\end{equation}
The photon-number conversion efficiency is therefore
\begin{equation}
\begin{aligned}
    &\eta_{\mathrm{ph}}=\frac{\lambda_o}{\lambda_s}\cdot\frac{|A_o(L)|^2}{|A_s|^2}=\frac{\lambda_o}{\lambda_s}J_1J_2P_{p1}P_{p2}|\mathcal I|^2,
\end{aligned}
\label{eq:eff_QFC}
\end{equation}
where the cascaded coupling integral $\mathcal I$ is also given by Eq.~\eqref{eq:cascaded_integral}.

Equations~\eqref{eq:eff} and~\eqref{eq:eff_QFC} show that maximizing the conversion efficiency of either cascaded SHG--SFG for third-harmonic generation or cascaded QFC amounts to maximizing the cascaded coupling integral $|\mathcal I|$ (Eq.~\eqref{eq:cascaded_integral}).

\subsection{Conventional two-section design}
Before we derive the optimum poling pattern for cascaded nonlinear processes, we first evaluate the conventional design that assigns the first length $L_1$ entirely to the first nonlinear process and the remaining length $L_2$ to the second. Its coupling factors are $(r_1,r_2)=(1,0)$ followed by $(0,1)$. Equation~\eqref{eq:cascaded_integral} gives the cascaded coupling integral 
\begin{equation}
 \mathcal I_{\mathrm{sections}}=L_1L_2
 \leq L^2/4,\qquad L=L_1+L_2.
 \label{eq:weaksections}
\end{equation}
The best split is therefore $L_1=L_2=L/2$ in this conventional design.

\subsection{Uniform beating poling pattern for simultaneous phase matching}
In addition to the conventional two-section design where two nonlinear processes are completely separated, we can also design a poling pattern that accommodates both nonlinear processes at the same time along the whole waveguide:
\begin{equation}
 s(z)=\operatorname{sgn}[\sin(K_1z)+\sin(K_2z)].
 \label{eq:uniformmask}
\end{equation}
This pattern follows the interference, or beating, of the two spatial tones. Its local Fourier spectrum contains components at both $K_1$ and $K_2$, each with complex Fourier amplitude $4/\pi^2$, compared with $2/\pi$ for a dedicated periodic grating. The component at $K_1$ compensates the mismatch of the first nonlinear interaction, while the component at $K_2$ compensates that of the second. Both interactions can therefore accumulate coherently throughout the same device.

The Fourier amplitude at each required wavevector is smaller than that of a dedicated periodic grating. The normalized coupling factors are therefore
\begin{equation}
 r_1=r_2=\frac{4/\pi^2}{2/\pi}=\frac{2}{\pi}.
 \label{eq:uniformcouplings}
\end{equation}
Substituting these constant coupling factors into Eq.~\eqref{eq:cascaded_integral} gives 
\begin{equation}
    \mathcal I_{\mathrm{beat}}=2L^2/\pi^2.
\end{equation}
Although both nonlinear processes use the full waveguide length, the reduction in the coupling factors outweighs the benefit of using the full waveguide length. Therefore, the final output is smaller than that of the conventional two-section design:
\begin{equation}
 \frac{\eta_{\mathrm{beat}}}{\eta_{\mathrm{sections}}}
 =\frac{64}{\pi^4}<1.
 \label{eq:uniformgain}
\end{equation}

This result follows from the buildup of the intermediate field. Near the input, little intermediate field is available, so the second interaction contributes weakly to the overall output. Increasing the second QPM coupling factor near the input reduces the first coupling factor available for intermediate-field generation. Near the exit, the intermediate field is already present and can be converted directly, whereas additional generation has little distance left to contribute to the overall output. Efficient conversion therefore favors the first interaction near the input and the second near the output. The conventional two-section design and the uniform beating poling pattern represent two extremes in the spatial allocation of nonlinear coupling. The former confines the interactions to disjoint sections, and the latter weights them equally throughout the waveguide. Both fall short of the maximum cascaded conversion efficiency.

\section{Optimum poling pattern for cascaded nonlinear processes}
In this section, we derive the optimum poling pattern analytically to achieve the maximum efficiency for cascaded nonlinear processes at low power.
For the optimization, we define the upstream and downstream interaction integrals, respectively,
\begin{equation}
 U(z)=\int_0^z r_1(t)\,dt,\qquad D(z)=\int_z^L r_2(t)\,dt.
 \label{eq:remaining}
\end{equation}
Here $U(z)$ measures the intermediate-field buildup before $z$, while $D(z)$ measures the remaining QPM interaction length available to convert that field downstream. Their endpoint conditions are $U(0)=0$ and $D(L)=0$. The cascaded coupling integral can be written in either form,\begin{equation} \mathcal I=\int_0^L U(z)r_2(z)\,dz =\int_0^L D(z)r_1(z)\,dz. \label{eq:integral_ud}\end{equation}where the second form follows by reversing the order of integration. 
To see how the poling at $z$ affects $\mathcal I$, we consider a short section $[z,z+\Delta z]$ with the poling outside this section held fixed. Intermediate field generated within this short section can be converted downstream, contributing $D(z)\int_z^{z+\Delta z}r_1(t)\,dt$. Intermediate field generated upstream can be converted within this short section, contributing $U(z)\int_z^{z+\Delta z}r_2(t)\,dt$. 
To leading order, the contribution in this short section is therefore $\big[D(z)r_1+U(z)r_2\big]\Delta z$. 



We denote the largest weighted coupling factor by
\begin{equation}
\begin{aligned}
 H(D,U)&=\max_{s=\pm1}\{D r_1+U r_2\}\\
 &=\frac{\pi}{2}\max_{s=\pm1}\avg{s\big[D\sin(K_1z)+U\sin(K_2z)\big]}\\
 &=\frac{\pi}{2}\avg{\big|D\sin(K_1z)+U\sin(K_2z)\big|}.
\end{aligned}
\label{eq:support}
\end{equation}
We use the same local-averaging approximation as in Eq.~\eqref{eq:rdefinition}, holding the slowly varying weights $D(z)$ and $U(z)$ fixed across the averaging window.
Choosing the domain sign to match the sign of the bracket makes the integrand its absolute value. 
Physically, $H(D,U)$ is the maximum available cascaded coupling factor weighted by the remaining conversion length and the intermediate field already generated.
The resulting poling pattern is globally optimal because it attains the upper bound on $\mathcal I$ in Eq.~\eqref{eq:bound}.


To express the relative tone weights, we introduce polar coordinates for the two nonnegative interaction integrals:
\[
\begin{aligned}
 D(z)&=\rho(z)\cos\theta(z),\\
 U(z)&=\rho(z)\sin\theta(z),\\
 \rho(z)&=\sqrt{D(z)^2+U(z)^2}.
\end{aligned}
\]
For $\rho>0$, multiplying the argument of the sign function by $\rho$ leaves the binary poling pattern $s(z)$ unchanged. Thus the common magnitude cancels from $\operatorname{sgn}[D\sin(K_1z)+U\sin(K_2z)]$, and the poling pattern depends only on the ratio $U(z)/D(z)$. The optimum poling pattern is therefore
\begin{equation}
\begin{aligned}s(z)=\operatorname{sgn}\!\big[&\cos\theta(z)\sin(K_1z)+\sin\theta(z)\sin(K_2z)\big].\end{aligned}
 \label{eq:mask}\end{equation}
with $\tan\theta(z)=U(z)/D(z)$.
The angle $\theta$ sets the relative weights of the two grating tones in the binary poling pattern. It increases from $\theta=0$ at the entrance ($z=0$) to $\theta=\pi/2$ at the exit ($z=L$), where the pattern provides QPM for the first and second nonlinear processes, respectively. The effective coupling factors $r_1$ and $r_2$ are obtained from the local Fourier coefficients of the resulting poling pattern $s(z)$ at $K_1$ and $K_2$, respectively [Eq.~\eqref{eq:rdefinition}].


\begin{figure*}[t]
 \centering\includegraphics[width=0.95\textwidth]{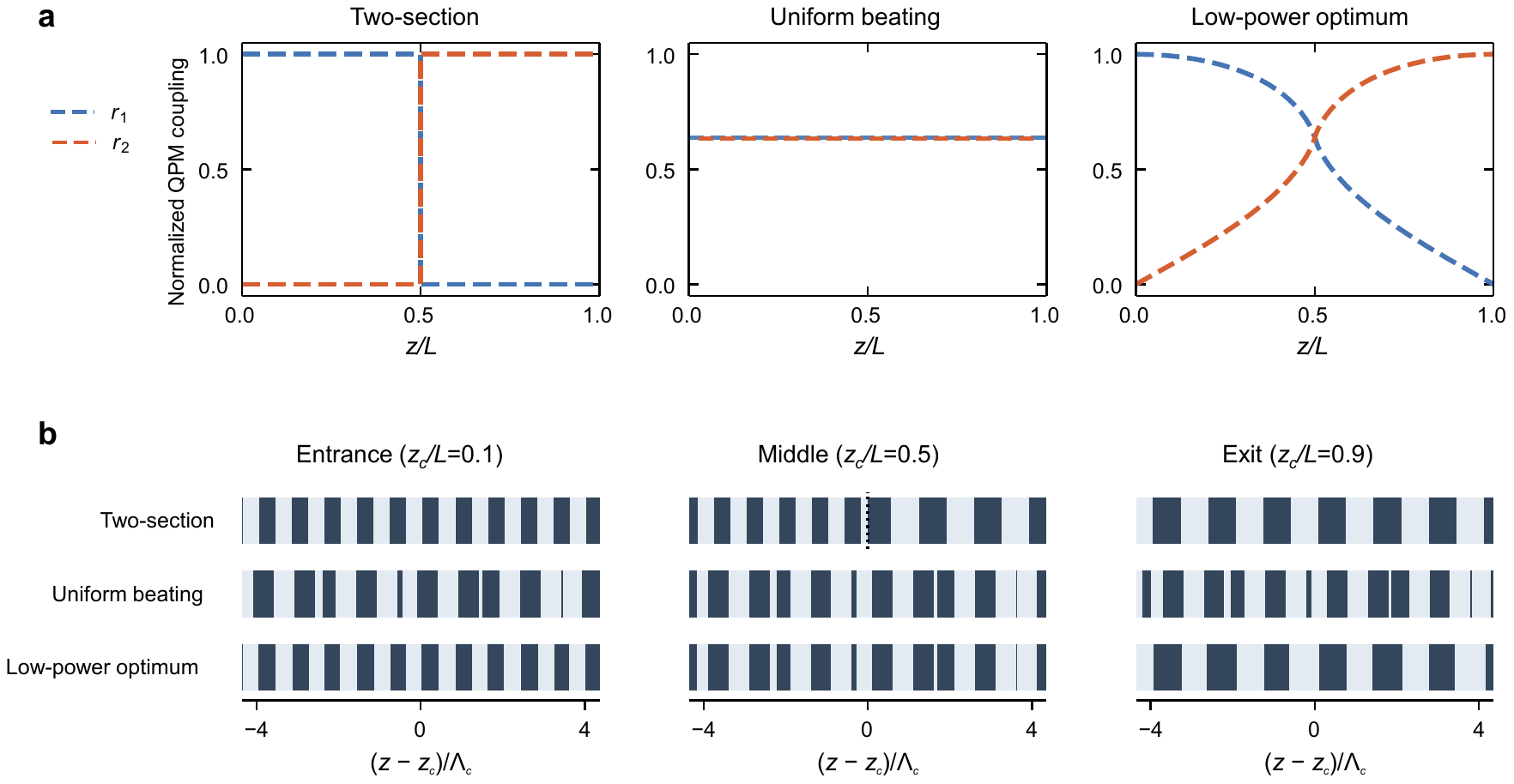}
 \caption{Effective coupling and binary poling layouts. (a) Coupling factors $r_1,r_2$ for equal-length two-section, uniform beating, and the low-power optimum designs, normalized to dedicated periodic QPM. The uniform-beating curves coincide at $2/\pi$. (b) Positive (dark) and negative (light) domain orientations at $z_c/L=0.1$, 0.5, and 0.9.}
 \label{fig:one}
\end{figure*}
The optimum poling pattern in Eq.~\eqref{eq:mask} is specified by the weighting angle $\theta(z)$. By homogeneity of $H$, the normalized weighted coupling $h(\theta)$ depends only on $\theta$:
\begin{equation}
\begin{aligned}
h(\theta)&\equiv\frac{H(D,U)}{\rho}\\
&=\frac{\pi}{2}\Big\langle\Big|\cos\theta\,\sin(K_1z)+\sin\theta\,\sin(K_2z)\Big|\Big\rangle\\
&=r_1\cos\theta+r_2\sin\theta.
\label{eq:norm_h}
\end{aligned}
\end{equation}

Under the independent-phase approximation, valid for incommensurate $K_1/K_2$ with many beat periods per averaging window, $K_1z$ and $K_2z$ inside the local average can be replaced by independent phases $\phi_1,\phi_2$ uniformly distributed on $[0,2\pi)$. The average in Eq.~\eqref{eq:norm_h} then has a closed form, and the coupling factors of the pattern in Eq.~\eqref{eq:mask} are
\begin{equation}
\begin{aligned}
&\theta\leq\pi/4:\\
&\quad r_1=\frac{2}{\pi}E(\tan^2\theta),\\
&\quad r_2=\frac{2}{\pi}\cot\theta\,\big[E(\tan^2\theta)-(1-\tan^2\theta)K(\tan^2\theta)\big],\\
&\theta>\pi/4:\\
&\quad r_1=\frac{2}{\pi}\tan\theta\,\big[E(\cot^2\theta)-(1-\cot^2\theta)K(\cot^2\theta)\big],\\
&\quad r_2=\frac{2}{\pi}E(\cot^2\theta),
\end{aligned}
\label{eq:elliptic}
\end{equation}
where $K(m)$ and $E(m)$ are complete elliptic integrals of the first and second kind with parameter $m$. 

With the optimum poling pattern, $H(D,U)$ is constant along the waveguide (see Appendix~\ref{app:constant_h}). Therefore, we can evaluate $H(D,U)$ at the waveguide output $z=L$:
\begin{equation}
    H(D,U)=U(L)\equiv F.
\end{equation}
The evolution of the angle $\theta(z)$ along the waveguide is then governed by (see Appendix~\ref{app:angle_evolution}) 
\begin{equation}
 \frac{dz}{d\theta}=\frac{F}{[h(\theta)]^2}.
 \label{eq:envelopeevolution}
\end{equation}

We define the cumulative length function $Q(\theta)$ and the dimensionless normalization constant $N=Q(\pi/2)$
\begin{equation}
\begin{gathered}
 Q(\theta)=\int_0^\theta\frac{d\vartheta}{[h(\vartheta)]^2},\qquad N=Q(\pi/2).
\end{gathered}
 \label{eq:profile}
\end{equation}
Integrating Eq.~\eqref{eq:envelopeevolution} from $\theta(0)=0$ gives $z=FQ(\theta(z))$. At the exit, $\theta(L)=\pi/2$, so $L=FQ(\pi/2)=FN$. Hence, we can obtain
\begin{equation}
\begin{aligned}
    &Q(\theta(z))=\frac{N\cdot z}{L}\\
    &\theta(z)=Q^{-1}\!\left(N\frac{z}{L}\right).\label{eq:thetaz}    
\end{aligned}
\end{equation}

With $\theta(z)$ determined, the design procedure is as follows. Given the mismatch wavevectors $K_1$ and $K_2$ and the waveguide length $L$, first evaluate $h(\theta)$ from Eqs.~\eqref{eq:norm_h} and \eqref{eq:elliptic}. Then integrate $[h(\theta)]^{-2}$ to obtain $Q(\theta)$ and its normalization $N$ [Eq.~\eqref{eq:profile}], tabulate $Q(\theta)/N$ on $0\leq\theta\leq\pi/2$, and invert this strictly increasing function at each fractional position $z/L$ to obtain $\theta(z)$ [Eq.~\eqref{eq:thetaz}]. Finally, insert $\theta(z)$ and the given wavevector ($K_1$ and $K_2$) into Eq.~\eqref{eq:mask} to obtain the binary poling pattern. 


Next, we show that this QPM design (Eq.~\eqref{eq:mask} - \eqref{eq:thetaz}) is optimum in the low-power, slow-varying regime.
We consider an arbitrary QPM pattern whose the intermediate field buildup at the exit is denoted by $u_f=U(L)$. 
Integrating the local bound on the weighted coupling factors along the waveguide gives
\begin{equation}
 \mathcal I\leq L u_f-\frac{N}{2}u_f^2\leq\frac{L^2}{2N}.
 \label{eq:bound}
\end{equation}
The optimized QPM design attains the first inequality by maximizing the weighted coupling at each position. It also gives $u_f=F=L/N$, the value that maximizes the quadratic in Eq.~\eqref{eq:bound}. It therefore attains the global upper bound on $\mathcal I$. Appendix~\ref{app:global_bound} derives the bound and verifies both equality conditions.

Evaluating Eq.~\eqref{eq:profile} with Eq.~\eqref{eq:elliptic} gives $N\approx1.7176$ and $\mathcal I_{\mathrm{opt}}/L^2=1/(2N)\approx0.2911$. The optimized output therefore exceeds that of two-section design by
\begin{equation}
\frac{\eta_{\mathrm{opt}}}{\eta_{\mathrm{sections}}}=\frac{4}{N^2}\approx1.356.
\label{eq:gains}
\end{equation}
The optimized QPM design builds the intermediate field near the input, overlaps the interactions through the center, and devotes the final region to output generation. 

The QPM coupling factors ($r_1$ and $r_2$) for two-section, uniform beating, and optimum designs are shown in Fig.~\ref{fig:one}a. The conventional two-section design shows the abrupt change between the two QPM sections; the uniform beating design shows the constant coupling factors for both nonlinear processes across the waveguide; the optimum design shows continuously varying coupling factors.
Fig.~\ref{fig:one}b shows the actual binary QPM patterns in three common-scale windows centered at locations $z_c/L=0.1$, 0.5, and 0.9, with local displacement normalized as $(z-z_c)/\Lambda_c$.
The optimum design shows similar QPM patterns as the conventional two-section design at the entrance and exit, to maximize the generation and conversion of the intermediate field. While in the middle, the optimum design shows similar QPM patterns with the uniform beating design, as the generation and conversion of the intermediate field are equally important.

\section{Limits imposed by finite poling domains}\label{sec:poling}
The analytical optimum is formulated in terms of locally averaged QPM factors. A fabricated device must instead realize a finite sequence of positive and negative domains, with a lower bound on every domain width. The relevant question is therefore how much conversion survives when this geometric constraint is applied to the optimum pattern.

For a finite device we allow constant phase offsets in the two grating tones:
\begin{equation}
\begin{aligned}
s(z)=\operatorname{sgn}\!\big[&\cos\theta(z)\sin(K_1z+\phi_1)\\
 &+\sin\theta(z)\sin(K_2z+\phi_2)\big].
\end{aligned}\label{eq:finitephases}
\end{equation}
The offsets leave the independent-phase averaged optimum unchanged, but shift individual domain walls relative to the envelope and device ends. They can therefore change the narrowest domains and the effect of enforcing a width limit. The binary geometry comparison in Fig.~\ref{fig:one} uses zero phases; the domain-width map below optimizes both offsets for each period ratio and width constraint.

Let $z_m$ denote consecutive domain boundaries, including the device ends. Fabrication imposes
\begin{equation}
 z_{m+1}-z_m\geq d_{\min}.
\end{equation}
We first locate the boundaries of Eq.~\eqref{eq:finitephases} with the analytical $\theta(z)$, and then repeatedly merge the shortest subthreshold domain into its neighbors until all widths satisfy this condition. The weak-conversion integral is then evaluated exactly for the resulting binary pattern (Appendix~\ref{app:finite}).

\begin{figure}[t]
\centering\includegraphics[width=\columnwidth]{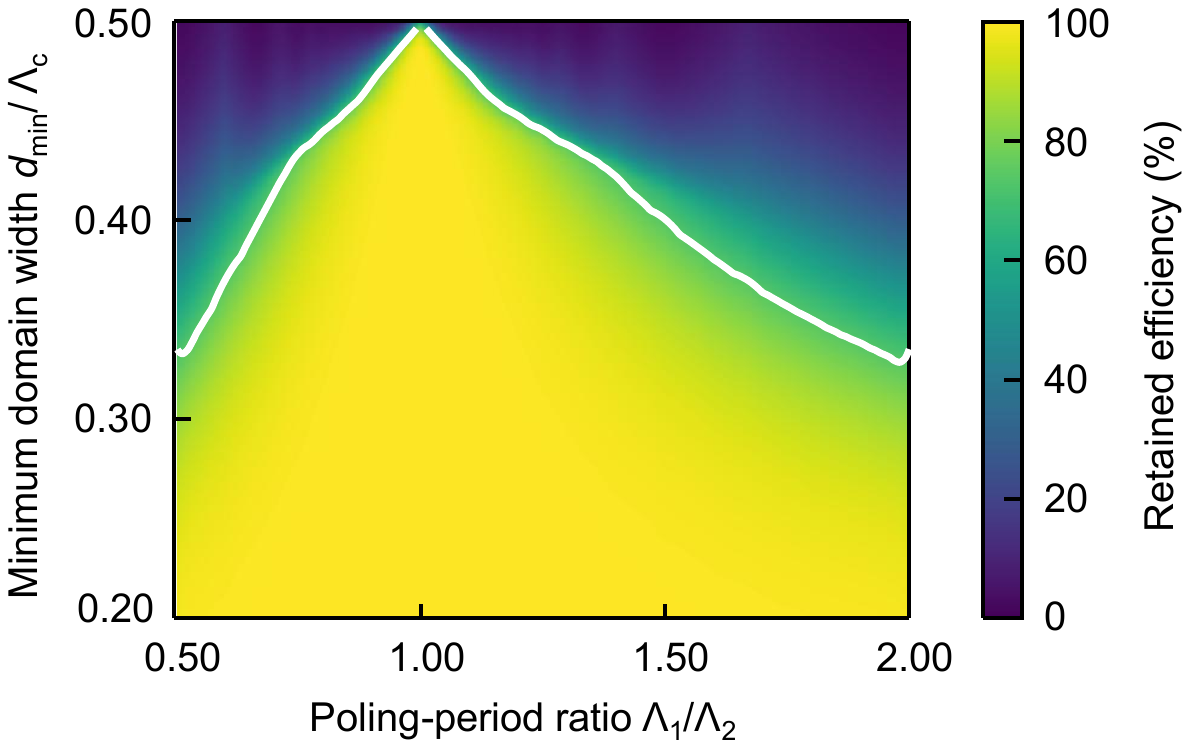}
\caption{Retained weak-conversion efficiency versus original QPM period ratio $\Lambda_1/\Lambda_2$ and relative minimum domain width $d_{\min}/\Lambda_c$ for waveguide length $L/\Lambda_c=3000$. The white dashed line marks the two-section design.}
\label{fig:three}\end{figure}

We use domain widths relative to the average poling period
\begin{equation}
 \Lambda_c=\frac{4\pi}{K_1+K_2}
 =\frac{2\Lambda_1\Lambda_2}{\Lambda_1+\Lambda_2},
 \qquad \Lambda_j=\frac{2\pi}{K_j}.
\end{equation}
The retained output is $|\mathcal I_{\mathrm{finite}}(d_{\min})/\mathcal I_{\mathrm{finite}}(0)|^2$. The efficiencies $J_1,J_2$ cancel from this ratio. 
Figure~\ref{fig:three} shows the retained efficiency over $0.5\leq\Lambda_1/\Lambda_2\leq2$ and $0.2\leq d_{\min}/\Lambda_c\leq0.5$ for a wavelength length of $L/\Lambda_c=3000$.
At $d_{\min}/\Lambda_c=0.2$, retention remains above 98\% over the sampled period ratios, showing the design robustness. The relative efficiency of the conventional two-section design is marked as white dashed line for reference.


\section{Complete cascaded quantum frequency conversion}\label{sec:complete}
The low-power optimum maximizes the ordered generation and conversion of the intermediate field. At high conversion, the same interactions also deplete the input and drive backconversion. We therefore retain the power-normalized amplitudes $A_s,A_i,A_o$, pump powers $P_{p1},P_{p2}$, efficiencies $J_1,J_2$, and local QPM factors $r_1,r_2$ introduced in Sec.~II, but include both directions of each interaction. After absorbing constant phases into the amplitudes, the lossless equations are
\begin{equation}
\begin{aligned}
 \frac{dA_s}{dz}&=-\frac{\lambda_i}{\lambda_s}\sqrt{J_1P_{p1}}\,r_1 A_i,\\
 \frac{dA_i}{dz}&=\sqrt{J_1P_{p1}}\,r_1 A_s
 -\frac{\lambda_o}{\lambda_i}\sqrt{J_2P_{p2}}\,r_2 A_o,\\
 \frac{dA_o}{dz}&=\sqrt{J_2P_{p2}}\,r_2 A_i.
\end{aligned}\label{eq:full}
\end{equation}
Here $A_i(0)=A_o(0)=0$, the pumps remain undepleted, and $\lambda_s,\lambda_i,\lambda_o$ are the signal-channel wavelengths. The wavelength factors follow from photon-flux conservation with power-normalized fields:
\begin{equation}
 \lambda_s|A_s|^2+\lambda_i|A_i|^2+\lambda_o|A_o|^2
 =\lambda_s|A_s(0)|^2.
\end{equation}
Thus $\eta_{\mathrm{ph}}=(\lambda_o/\lambda_s)|A_o(L)/A_s(0)|^2$, consistently with Eq.~\eqref{eq:eff_QFC}. Neglecting the reverse terms recovers Eq.~\eqref{eq:qfcweakfields}. Complete conversion requires $A_s(L)=A_i(L)=0$; we minimize $P_{p1}+P_{p2}$ at fixed $L$.

We consider $L=1$~cm, $\lambda_s=397$~nm, and pumps at 1550 and 1935~nm. Energy conservation fixes $\lambda_i\approx533.7$~nm and $\lambda_o\approx737.0$~nm. We calculate the nonlinear overlaps of the X-cut thin-film lithium niobate ridge waveguides described in Appendix~\ref{app:nonlinear_overlaps}. Related thin-film lithium niobate waveguides have been used for SHG, QFC, and parametric amplification~\cite{Chen2024adapted, briggs2026noise, kao2026efficient}. With power normalization and the first-order QPM factor included, the type-0 $d_{33}$ model gives
\begin{equation}
 J_1\simeq364.3,\qquad J_2\simeq127.5
 \quad (\mathrm{W}^{-1}\mathrm{cm}^{-2}).\label{eq:simulated_J}
\end{equation}
Nonlinear-material dispersion is included through a Miller extrapolation from the measured $d_{33}=25.2$~pm/V for 1064-to-532-nm SHG~\cite{Shoji1997,Alford2001}.

For the two-section design, complete conversion in each section requires the individual pump powers
\begin{equation}
 P_{p1}=\frac{\pi^2\lambda_s}{4J_1\lambda_i L_1^2},\qquad
 P_{p2}=\frac{\pi^2\lambda_i}{4J_2\lambda_o L_2^2}.
 \label{eq:sectionpowers}
\end{equation}
For a fixed total length $L_1+L_2=L$, we define $P_{\mathrm{sections}}$ as the minimum total pump power $P_{p1}+P_{p2}$ required for complete cascaded conversion.
\begin{equation}
\begin{aligned}
 P_{\mathrm{sections}}
 &\equiv\min_{\substack{L_1+L_2=L}}(P_{p1}+P_{p2})\\
 &=\frac{\pi^2}{4L^2}
 \left[\left(\frac{\lambda_s}{J_1\lambda_i}\right)^{1/3}
 +\left(\frac{\lambda_i}{J_2\lambda_o}\right)^{1/3}\right]^3.
\end{aligned}\label{eq:sectionsfull}
\end{equation}
This minimum is attained when the two QPM section lengths satisfy
\begin{equation}
 \frac{L_1}{L_2}=\left(\frac{J_2\lambda_s\lambda_o}{J_1\lambda_i^2}\right)^{1/3}.
 \label{eq:optimalsectionratio}
\end{equation}

For uniform beating, $r_1=r_2=2/\pi$ and complete transfer requires equal photon-normalized coupling rates. The required powers are
\begin{equation}
\begin{aligned}
P_{p1}&=\frac{\pi^4\lambda_s}{8J_1\lambda_i L^2},\qquad
P_{p2}=\frac{\pi^4\lambda_i}{8J_2\lambda_o L^2},\\
P_{\mathrm{beat}}&=P_{p1}+P_{p2}=\frac{\pi^4}{8L^2}\left(\frac{\lambda_s}{J_1\lambda_i}+\frac{\lambda_i}{J_2\lambda_o}\right).
\end{aligned}
\label{eq:beatpowers}
\end{equation}

\begin{figure}[t]
\centering\includegraphics[width=\columnwidth]{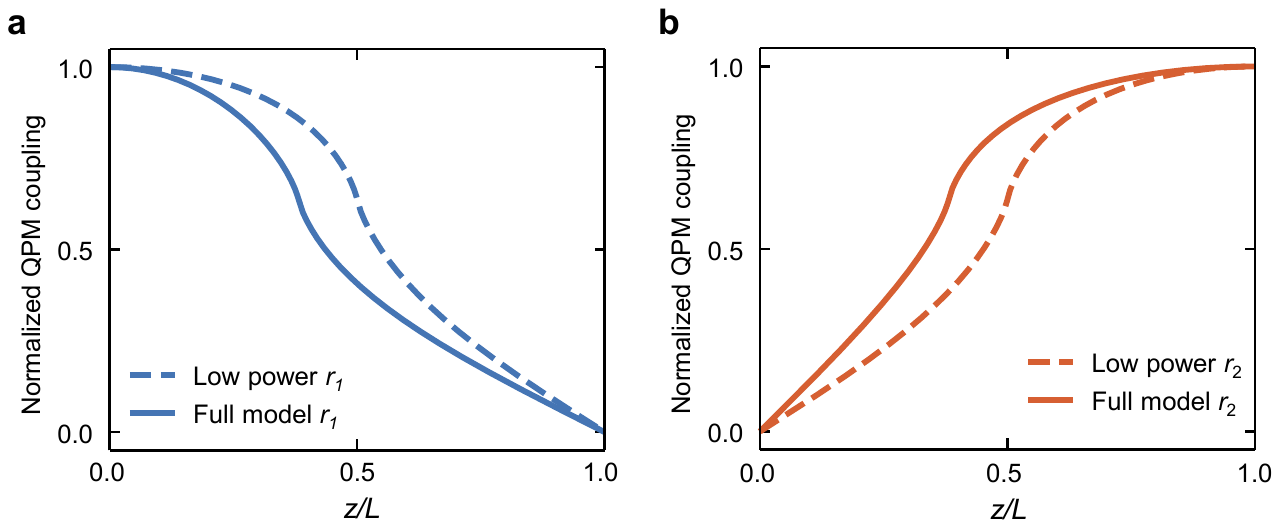}
\caption{Low-power (dashed) and complete-transfer (solid) effective coupling profiles for the (a) first and (b) second nonlinear processes.
}
\label{fig:two}\end{figure}

For the simulated efficiencies, optimized two-section QPM requires $P_{\mathrm{sections}}\approx70.2$~mW and uniform beating requires 94.0~mW. Keeping the analytical weak-conversion envelope fixed and adjusting only the two pump powers gives complete conversion at 62.80~mW, 10.6\% below the two-section design. The low-power design therefore retains a clear advantage at complete conversion, although it is no longer optimal. Optimizing the envelope and pump allocation together with Pontryagin's maximum principle~\cite{Boscain2021} (Appendix~\ref{app:pmp}) gives 58.17~mW (23.47~mW at 1550~nm and 34.69~mW at 1935~nm); the analytical design requires 7.97\% more. Figure~\ref{fig:two} compares the two coupling profiles. 

\section{Discussion and outlook}
We have derived a spatially weighted QPM design for cascaded second-order frequency conversion. The design follows the evolution of the intermediate field: QPM coupling is concentrated in the first nonlinear interaction near the entrance and transferred gradually to the second nonlinear interaction toward the exit. This preserves the ordering required by a cascade while allowing both nonlinear processes to act over overlapping regions of the waveguide.

In the lossless weak-conversion regime, the analytical envelope gives 35.6\% higher output than optimized separate sections and attains the efficiency bound of the slow-varying binary-poling model for any pair of nonlinear efficiencies. With a minimum domain width of $0.2\Lambda_c$, the phase-optimized patterns retain at least 98\% of the unconstrained output for $L/\Lambda_c=3000$. At complete conversion in the simulated waveguide, it requires 10.6\% less pump power than optimized sections and lies within 8\% of a numerically optimized design, making it a practical starting point when depletion matters. 

In a fabricated waveguide, film-thickness variations can shift the local phase mismatches away from the design values $K_1$ and $K_2$~\cite{Kuo2022noncritical}. Measuring the waveguide geometry before poling provides a way to account for these variations when specifying the grating~\cite{Xin2025wafer}. The operating pump power can also matter. Simulations of SHG show that photorefraction can introduce a longitudinal mismatch that can be compensated by varying the local poling period~\cite{Wang2025photorefractive}.


\begin{acknowledgments}
This work was supported in part by the Defense Advanced Research Projects Agency (DARPA) HARQ program HR0011-26-9-E107, Office of Naval Research (N00014-25-1-2130), Air Force Office of Scientific Research (FA9550-24-1-0119), U.S. Department of Energy (Field Work Proposal ERKJ432). L.F. acknowledges the support from Coherent/II-VI Foundation and Sloan Fellowship. 

GPT-6 Astra is used for the optimality proof of the QPM design and the numerical calculation of nonlinear efficiencies. Opus 5.5 is used for cross-check the proof.
\end{acknowledgments}

\section*{Data Availability}
Source Data will be provided upon reasonble request.

\appendix
\section{Constancy of the weighted coupling along the optimum poling pattern}
\label{app:constant_h}
For a given poling pattern, let $r_1$ and $r_2$ denote its local coupling factors and let $r_1^\star$ and $r_2^\star$ denote the coupling factors that maximize $Dr_1+Ur_2$ in Eq.~\eqref{eq:support} at the current weights $D(z)$ and $U(z)$. Where $H$ is differentiable, the envelope theorem gives $\partial H/\partial D=r_1^\star$ and $\partial H/\partial U=r_2^\star$. The actual coupling factors set the evolution of the weights through $D'=-r_2$ and $U'=r_1$, where primes denote derivatives with respect to $z$. The chain rule gives
\[
\begin{aligned}
\frac{d}{dz}H(D(z),U(z))
  &= \frac{\partial H}{\partial D}D'
    + \frac{\partial H}{\partial U}U' \\
  &= r_2^\star r_1-r_1^\star r_2.
\end{aligned}
\]
Along the optimum poling pattern, $(r_1,r_2)=(r_1^\star,r_2^\star)$, so $dH/dz=0$. The same cancellation can occur when $(r_1,r_2)$ is proportional to $(r_1^\star,r_2^\star)$, which makes constancy of $H$ insufficient to establish optimality. Along this optimum pattern, $D(L)=0$ at the exit and $H(0,U)=U$ for $U\geq 0$, so
\[
H(D(z),U(z))=H(0,U(L))=U(L)\equiv F.
\]
Here $F$ is the final intermediate-field buildup factor.

\section{Evolution of the weighting angle}
\label{app:angle_evolution}
Write $D=\rho\cos\theta$ and $U=\rho\sin\theta$, with primes denoting derivatives with respect to $z$. Differentiating both expressions and forming $DU'-UD'$ cancels the terms proportional to $\rho'$, giving
\begin{equation}
 \rho^2\frac{d\theta}{dz}
 = DU'-UD'
 = Dr_1+Ur_2,
\end{equation}
where the last equality uses $U'=r_1$ and $D'=-r_2$. Along the optimum poling pattern, $Dr_1+Ur_2=H(D,U)=F$ by Eq.~\eqref{eq:support} and Appendix~\ref{app:constant_h}. Since $\rho=F/h(\theta)$,
\begin{equation}
\begin{aligned}
 \frac{d\theta}{dz}
 &=\frac{DU'-UD'}{D^2+U^2}
 =\frac{Dr_1+Ur_2}{D^2+U^2}\\
 &=\frac{F}{\rho^2}
 =\frac{[h(\theta)]^2}{F}.
\end{aligned}
\end{equation}
For a nontrivial optimum, $F>0$ and $h(\theta)>0$, so $\theta(z)$ is strictly increasing and can be inverted:
\begin{equation}
 \frac{dz}{d\theta}=\frac{F}{[h(\theta)]^2}.
\end{equation}

\section{Global optimality of the weighting envelope}
\label{app:global_bound}
This appendix proves Eq.~\eqref{eq:bound}: no poling pattern can give a cascaded coupling integral larger than $L^2/(2N)$, and the analytical envelope reaches this value. We work in the lossless, weak-conversion model with independent phase averaging and consider patterns whose coupling factors are nonnegative and phase aligned. The proof has four steps. We first bound the integrand at each position, then integrate this bound along the device, then optimize the result over the final intermediate-field buildup, and finally show that the analytical pattern turns every inequality into an equality.

\textit{Starting point.} For an arbitrary poling pattern with coupling factors $r_1(z)$ and $r_2(z)$, Eq.~\eqref{eq:integral_ud} gives
\begin{equation}
\mathcal I=\int_0^L U(z)\,r_2(z)\,dz,\qquad U(z)=\int_0^z r_1(t)\,dt.
\label{eq:C_start}
\end{equation}
The integrand is the intermediate field already generated, $U(z)$, multiplied by the local strength $r_2(z)$ with which it is converted to the output. Because a single binary pattern supplies both coupling factors, increasing $r_2$ at a given position generally reduces $r_1$ there. The function $H$ in Eq.~\eqref{eq:support} quantifies this trade-off: for any nonnegative weights $a$ and $u$ and any pair $(r_1,r_2)$ produced by a binary pattern,
\begin{equation}
a\,r_1+u\,r_2\leq H(a,u).
\label{eq:C_tradeoff}
\end{equation}

\textit{Step 1: a local bound with a common constant.} Let $u_f=U(L)$ denote the final buildup of the pattern under consideration. Since $U'=r_1\geq0$, $U(z)$ increases monotonically from $0$ to $u_f$. If $u_f=0$, then $U\equiv0$ and $\mathcal I=0$, so we assume $u_f>0$.

For each buildup level $0\leq u\leq u_f$, we choose a comparison weight $a=D_{u_f}(u)\geq0$ defined by
\begin{equation}
H\big(D_{u_f}(u),u\big)=u_f,
\label{eq:C_level}
\end{equation}
taking the largest solution if more than one exists. In words, $D_{u_f}(u)$ is the downstream weight that, combined with the buildup $u$, gives the same maximum weighted coupling as the device exit, where $(D,U)=(0,u_f)$ and $H(0,u_f)=u_f$. A solution always exists. At $a=0$, $H(0,u)=u\leq u_f$. At $a=u_f$, the dedicated first-process grating $(r_1,r_2)=(1,0)$ shows that $H(u_f,u)\geq u_f$. Between these values, $H$ varies continuously with $a$, because changing $a$ by $\delta a$ changes every weighted sum by at most $|\delta a|$.

Setting $u=U(z)$ and $a=D_{u_f}(U(z))$ in Eq.~\eqref{eq:C_tradeoff} and using Eq.~\eqref{eq:C_level} gives, at every position,
\begin{equation}
U(z)\,r_2(z)\leq u_f-D_{u_f}\big(U(z)\big)\,r_1(z).
\label{eq:C_local}
\end{equation}
The right-hand side contains the same constant $u_f$ everywhere along the waveguide, which makes the bound straightforward to integrate.

\textit{Step 2: integration along the device.} Integrating Eq.~\eqref{eq:C_local} from $z=0$ to $z=L$ and changing variables with $du=dU=r_1\,dz$ in the second term gives
\begin{equation}
\mathcal I\leq L\,u_f-\int_0^{u_f}D_{u_f}(u)\,du.
\label{eq:C_integrated}
\end{equation}
The remaining integral depends only on $u_f$ and on the function $H$, not on the details of the pattern, and it has a simple geometric meaning. Consider the region $\mathcal R=\{(a,u):a\geq0,\ u\geq0,\ H(a,u)\leq u_f\}$ in a plane with the weight $a$ on the horizontal axis and the buildup $u$ on the vertical axis. Because $H$ does not decrease as $a$ increases, the horizontal slice of $\mathcal R$ at height $u$ extends from $a=0$ to $a=D_{u_f}(u)$. The integral in Eq.~\eqref{eq:C_integrated} is therefore the area of $\mathcal R$.

To compute this area, we introduce polar coordinates $a=\rho\cos\theta$ and $u=\rho\sin\theta$ in this plane, analogous to those used for $(D,U)$ in the main text. Here $\rho$ and $\theta$ are coordinates in the $(a,u)$ plane, not functions of $z$. The homogeneity of $H$ and the definition of $h$ in Eq.~\eqref{eq:norm_h} give $H(a,u)=\rho\,h(\theta)$, so the boundary of $\mathcal R$ lies at radius $\rho=u_f/h(\theta)$. Summing the areas $\tfrac12\rho^2\,d\theta$ of thin angular sectors from $\theta=0$ to $\pi/2$ gives
\begin{equation}
\int_0^{u_f}D_{u_f}(u)\,du=\frac{u_f^2}{2}\int_0^{\pi/2}\frac{d\theta}{[h(\theta)]^2}=\frac{N u_f^2}{2},
\label{eq:C_area}
\end{equation}
where $N$ is the normalization constant defined in Eq.~\eqref{eq:profile}.

\textit{Step 3: optimization over the final buildup.} Substituting Eq.~\eqref{eq:C_area} into Eq.~\eqref{eq:C_integrated} and completing the square gives
\begin{equation}
\begin{aligned}
\mathcal I&\leq L\,u_f-\frac{N u_f^2}{2}\\
&=\frac{L^2}{2N}-\frac{N}{2}\left(u_f-\frac{L}{N}\right)^2\leq\frac{L^2}{2N}.
\end{aligned}
\label{eq:C_bound}
\end{equation}
This result holds for every pattern in the model and reproduces Eq.~\eqref{eq:bound}. The first inequality becomes an equality only if Eq.~\eqref{eq:C_local} is an equality at every position, and the second only if $u_f=L/N$.

\textit{Step 4: the analytical pattern attains the bound.} The analytical pattern satisfies both equality conditions. Along this pattern, $H(D(z),U(z))=F=u_f$ (Appendix~\ref{app:constant_h}), so the comparison weight coincides with the actual downstream integral, $D_{u_f}(U(z))=D(z)$. Because the pattern in Eq.~\eqref{eq:mask} maximizes the weighted coupling at every position, Eq.~\eqref{eq:C_tradeoff} and therefore Eq.~\eqref{eq:C_local} hold with equality:
\begin{equation}
D(z)\,r_1(z)+U(z)\,r_2(z)=H\big(D(z),U(z)\big)=F.
\label{eq:C_equality}
\end{equation}
By Eq.~\eqref{eq:integral_ud}, each term on the left-hand side integrates to $\mathcal I$, so integrating Eq.~\eqref{eq:C_equality} over the device gives $2\mathcal I=FL$. Integrating $dz/d\theta=F/[h(\theta)]^2$ from $\theta=0$ to $\pi/2$ gives $L=FN$ (Appendix~\ref{app:angle_evolution}), so $u_f=F=L/N$, which satisfies the second equality condition. Combining these results,
\begin{equation}
\mathcal I_{\mathrm{opt}}=\frac{FL}{2}=\frac{L^2}{2N},
\end{equation}
which equals the upper bound in Eq.~\eqref{eq:C_bound}. The analytical envelope is therefore globally optimal within the lossless, slow-varying weak-conversion model.

\section{Exact evaluation of finite poling patterns}
\label{app:finite}
For a finite binary pattern $s(z)$, the weak-conversion integral is evaluated without local averaging,
\begin{equation}
\mathcal I_{\mathrm{finite}}=\left(\frac{\pi}{2}\right)^2\int_0^L s(z)e^{iK_2z}\int_0^z s(t)e^{iK_1t}\,dt\,dz,
\end{equation}
where the prefactor converts the raw Fourier coefficients to the periodic-QPM normalization. On each domain of constant sign both integrals are elementary, so $\mathcal I_{\mathrm{finite}}$ follows exactly from the domain boundaries, which are located by root finding on Eq.~\eqref{eq:finitephases}. The calculation retains all spatial harmonics of the binary pattern within the two-process model. 

\section{Waveguide model and nonlinear overlaps}
\label{app:nonlinear_overlaps}
The waveguide is an X-cut lithium-niobate ridge with 600-nm film thickness, 250-nm slab thickness, 2.2-$\mu$m top width, 70-degree sidewalls, and silica cladding. Light propagates along the crystal $Y$ axis, so the TE-like modes couple through $d_{33}$. Modes are computed using the extraordinary and ordinary indices ($\lambda$ in $\mu$m)
\begin{equation}
\begin{aligned}
n_e^2&=4.5820+\frac{0.099169}{\lambda^2-0.04443}-0.021950\lambda^2,\\
n_o^2&=4.9048+\frac{0.11768}{\lambda^2-0.04750}-0.027169\lambda^2,
\end{aligned}
\end{equation}
from Ref.~\cite{Smith1976}, and tabulated data for silica~\cite{Philipp1985}.

For each mode, the forward power is
\begin{equation}
 P_\mu=\frac12\operatorname{Re}\int (\mathbf E_\mu\times\mathbf H_\mu^*)\cdot\hat{\mathbf z}\,dS,
 \qquad \mathbf e_\mu=\mathbf E_\mu/\sqrt{P_\mu}.
\end{equation}
The DFG coefficient for $a\rightarrow b$ with pump $p$ is
\begin{equation}
 \label{eq:Jdfg}\sqrt{J_{a\rightarrow b}}=\frac{2}{\pi}\frac{\omega_b\epsilon_0}{2}
 \left|\int_{\mathrm{LN}}d_{33}(\omega_a,\omega_b,\omega_p)
 e_{b,x}^*e_{a,x}e_{p,x}^*\,dS\right|.
\end{equation}
This expression gives $J$ in $\mathrm{W}^{-1}\mathrm{m}^{-2}$. Integration includes the ridge and slab, excluding silica. 

No exact $d_{33}$ values for the target DFG triplets were identified in the checked references. We use the explicit extrapolation
\begin{equation}
 \label{eq:miller}d_{33}(\lambda_a,\lambda_b,\lambda_p)=d_{\rm ref}
 \frac{\chi_e(\lambda_a)\chi_e(\lambda_b)\chi_e(\lambda_p)}
 {\chi_e(532\,\mathrm{nm})[\chi_e(1064\,\mathrm{nm})]^2},
\end{equation}
where $\chi_e(\lambda)=n_e^2(\lambda)-1$ and $d_{\rm ref}=25.2$~pm/V from the congruent-LN SHG measurement in Table~6 of Ref.~\cite{Shoji1997}. The extraordinary-index model is the same as that used for the mode calculations. 

\section{Numerical optimization at complete conversion}
\label{app:pmp}
With photon-flux-normalized amplitudes $\mathbf u=(A_s,\sqrt{\lambda_i/\lambda_s}\,A_i,\sqrt{\lambda_o/\lambda_s}\,A_o)/A_s(0)$, Eq.~\eqref{eq:full} becomes a norm-preserving rotation, $d\mathbf u/d\tau=\boldsymbol\Omega\times\mathbf u$, with $\tau=z\sqrt{P_\Sigma}$, $P_{p1}=qP_\Sigma$, $P_{p2}=(1-q)P_\Sigma$, and $\boldsymbol\Omega=(c\,r_2,0,a\,r_1)$, where $a^2=qJ_1\lambda_i/\lambda_s$ and $c^2=(1-q)J_2\lambda_o/\lambda_i$. Complete conversion rotates $\mathbf u$ from $(1,0,0)$ to $(0,0,1)$, and the required total pump power is $(T/L)^2$, where $T$ is the scaled transfer length. Minimizing $T$ is a time-optimal control problem with the same admissible coupling set as in Sec.~III. Pontryagin's maximum principle selects at each position the binary pattern of Eq.~\eqref{eq:mask}, with a weighting angle set by the costate, and reduces the problem to shooting on the first transfer branch followed by a one-dimensional minimization over $q$, giving $q\approx0.404$. The endpoint is verified independently by piecewise exact rotations. 

\bibliography{references}
\end{document}